# Giant Optical Polarization Rotation Induced by Spin-Orbit Coupling in Polarons

Blai Casals[1], Rafael Cichelero[1], Pablo García Fernández[2], Javier Junquera[2], David Pesquera[1], Mariano Campoy-Quiles[1], Ingrid C. Infante[1,*], Florencio Sánchez[1], Josep Fontcuberta[1], Gervasi Herranz[1]

[1]Institut de Ciència de Materials de Barcelona (ICMAB-CSIC), Campus de la UAB, 08193 Bellaterra, Catalonia, Spain

[2]Departamento de Ciencias de la Tierra y Física de la Materia Condensada, Universidad de Cantabria, Cantabria Campus Internacional, Avenida de los Castros s/n, 39005 Santander, Spain

*We have uncovered a giant gyrotropic magneto-optical response for doped ferromagnetic manganite $La_{2/3}Ca_{1/3}MnO_3$ around the near room-temperature paramagnetic-to-ferromagnetic transition. At odds with current wisdom, where this response is usually assumed to be fundamentally fixed by the electronic band structure, we point to the presence of small polarons as the driving force for this unexpected phenomenon. We explain the observed properties by the intricate interplay of mobility, Jahn-Teller effect and spin-orbit coupling of small polarons. As magnetic polarons are ubiquitously inherent to many strongly correlated systems, our results provide an original, general pathway towards the generation of gigantic gyrotropic responses that can be harnessed for nonreciprocal devices that exploit the polarization of light.*

Polarons, first conceived by Landau [1], are quasiparticles formed by electrons that are bound to lattice deformations. Their physics is underpinned by electron-phonon interactions, which dominate transport in semiconductors, many poor metals and organics [2]. When the electronic bandwidth $W$ is sufficiently large Fröhlich-large polarons spread over many lattice sites with nearly free-electron propagation and slightly increased effective mass [2-4]. Yet, when the polaron binding energy $E_b$ is larger than half-bandwidth, i.e., $\lambda = 2E_b/W > 1$, the coupling to the lattice is so strong that electronic states are heavily dressed by phonons and electrons are self-trapped, forming small Holstein polarons [5-7]. In the latter, the transport rather than diffusive proceeds by thermally activated hopping.

Interestingly, thermal agitation or phonons are not the only ways to prompt polaron hopping; light provides an additional pathway. In particular, the absorption of photons may deliver the required energy to jump between sites –i.e., approximately $E_b$/2 or half the polaron binding energy–. In the presence of a magnetic field breaking time-reversal symmetry, the conductivity tensor acquires asymmetric non-diagonal terms, thus generating gyrotropic responses [8]; as a consequence, light with different circularly polarized states is absorbed differently, inducing rotation and ellipticity in the polarization of light [2,9-11]. Particularly remarkable, the strong coupling of Holstein polarons to the lattice implies vibronic states with energy scales close to the eV, thus conceivably enabling gyrotropic responses in the optical range [8]. So far, though, the specific contribution of small polarons to the gyrotropic effect has remained unsolved.

Here we disclose a unique gyrotropic response of small polarons in the visible range and uncover that Holstein-like hopping transport coupled to spin-orbit coupling leads to a gigantic magneto-optical response. To demonstrate this outstanding phenomenon, we selected optimally doped ferromagnetic manganites of composition $RE_{1-x}A_xMnO_3$ ($x \approx 0.3$) –where $RE$ is a rare earth and *A* an alkaline element–. Since a large part of the polaron binding energy $E_b$ can be identified with the Jahn-Teller energy $E_{JT}$ arising from strong electron-phonon interactions, the selected materials are ideal for test-bedding polaron optical responses. In these systems a colossal magnetoresistance (CMR) [12, 13] is observed around the paramagnetic-to-ferromagnetic transition. The ultimate reason is that electrons are self-trapped in narrow *d*-orbitals of Mn with $e_g$ symmetry forming magnetic small polarons that are sharply suppressed below the Curie temperature $T_C$. Two different responses –sketched in Figure 1– reveal the effect of polarons on the optical properties of CMR manganites [14,15]. First, at temperatures below $T_C$ a magneto-optical (MO) signal develops, which is proportional to the magnetization and reflects its typical temperature dependence (Figure 1c). This term is gyrotropic and implies the conversion between the orthogonal s- and p- components of polarization, so that polarized light undergoes rotation and ellipticity [10,11] (Figure 1a, top). Secondly, an additional contribution is revealed by a steep change of the optical reflectance with field, henceforth referenced as magneto-reflectance [14,15] (MR, Figures 1a and 1b). The MR exhibits a cusp-like dependence centered on the ferromagnetic transition, signaling the temperature interval in which polarons exist (Figure 1c). This observation, endorsed by multiple optical conductivity spectroscopy experiments [16-18], backs up MR as the optical counterpart of CMR. The MR is non-gyrotropic and, for the isotropic non-depolarizing case, entails a change in the intensity of light, without mixing its polarization components (Figure 1a, bottom). Summing up, MO and MR can be used as proxies, respectively, for magnetism and polaron signatures and their coexistence is translated into anomalous magneto-optical loops, as depicted pictorially in Figure 1b.

To study the interplay between polarons and gyrotropic responses we chose the La$_{2/3}$Ca$_{1/3}$MnO$_3$ manganite. The reason is that he bandwidth $W$ of the $e_g$ states is narrow in this material [19] and thus the adimensional electron-phonon coupling constant is large ($\lambda = 2E_b/W > 1$) and therefore small polaron transport is prevalent [16,20,21]. For that purpose, La$_{2/3}$Ca$_{1/3}$MnO$_3$ thin films were grown by rf sputtering on (110)-oriented SrTiO$_3$ single crystals. During the deposition, the substrate was held at a temperature of 800 ºC and a pressure of 330 mTorr, with an O$_2$/Ar pressure ratio of 1/4. After growth, the samples were annealed in situ at 800 °C for 1 h in an O$_2$ atmosphere at 350 Torr. Although the main text focuses the discussion on the film with thickness $t = 80$ nm, our study comprised the analysis of more samples (see the Supplementary Information), in which the thickness ranged within the interval $17$ nm $< t < 93$ nm, for films also grown on (110)-oriented SrTiO$_3$ substrates.

The optical characterization of the samples was done using a transverse configuration and using the null-ellipsometry method to extract the magneto-optical responses [22]. The transverse Kerr signal ($\delta_K = \Delta I_p / I_p$), generally a complex magnitude [10,11], is defined by the relative change with magnetic field of the intensity $I_p$ of p-polarized light –i.e. light with its polarization contained in the plane of incidence– upon reflection on a magnetic surface. As we have discussed previously [14,15], the different parity symmetries of $\delta_K$ with respect to the applied fields –even vs odd– can be exploited to decompose the hysteretic $\delta_K$ loops into gyrotropic MO $\Im(\delta_K)_{MO}$ and nongyrotropic MR $\Im(\delta_K)_{MR}$ terms, respectively. In Figure 2a (top panel) we show the raw data of the imaginary part of the transverse Kerr response $\Im(\delta_K)$ measured at a temperature $T \approx 256$ K when cycling the magnetic field up to $\approx \pm 9$ kOe. The decomposition of $\Im(\delta_K)$ into the $\Im(\delta_K)_{MO}$ and $\Im(\delta_K)_{MR}$ terms is shown in the middle and bottom panels of Figure 2a, respectively.

The complementarity of MO and MR is the stepping stone to understand how polarons affect the gyrotropic activity. A first insight is given by the wavelength dependence of $\Im(\delta_K)_{MO}$ and $\Im(\delta_K)_{MR}$ at the highest applied field ($H \sim 15$ kOe). As it turns out, the spectral responses shown in Figure 2b are dominated by two distinctive lobes located, respectively, at long ($\lambda_{ph} >\sim$ 550 nm, $E_{ph} < \sim 2.3$ eV) and short ($\lambda_{ph} \sim < 550$ nm, $E_{ph} > \sim 2.3$ eV) wavelengths, hinting at two broad electronic transitions excited by photons. Variable angle spectroscopic ellipsometry corroborate this view, as they show that the imaginary part of the dielectric function ($\varepsilon_2$) can be also described by two transitions (Figure 2c, and Supplementary Information). Indeed, our results are consistent with a large number of spectroscopy studies in CMR manganites showing two main bands extending around $E_{ph} \sim 1-2$ eV and $E_{ph} \sim 3-5$ eV, respectively [16-18, 23-25]. Such spectral features have received different interpretations in terms of either intrasite [17,23] or intersite transitions between $e_g$ states for bands at $E_{ph} \sim 1-2$ eV as well as intraatomic charge transfer excitations for optical bands at $E_{ph} \sim 4-4.5$ eV [25]. We make a special mention here of the interpretation put forward by Quijada et al. [16], who claimed the polaronic character of the observed spectral features [16,23]. They assigned the lower-energy bands at $E_{ph} < \sim 2$ eV to intersite $e_g - e_g$ photoinduced transitions of small polarons that hop between neighboring Mn sites without changing their spin, while optical bands at $E_{ph} \sim 3$ eV were assigned to intersite $e_g - e_g$ transitions in which the spin state is changed [16]. In the following we show that this spin-dependent photoinduced polaron-hopping, which is drawn pictorially in Figure 2d, can explain the outstanding gyrotropic response of La$_{2/3}$Ca$_{1/3}$MnO$_3$. In view of this, we define two spectral regions (see Figures 2b and 2c) that select the photon wavelengths at which the spin is either preserved (*SP*) or flipped (*SF*) during the photoinduced transitions of small polarons.

Figure 3a represents the curves of $\Im(\delta_K)_{MO}$ in the range of temperatures of $230-290$ K (the Curie temperature of La$_{2/3}$Ca$_{1/3}$MnO$_3$ is $T_C \approx 265$ K). Examination of this Figure reveals that a bump in $\Im(\delta_K)_{MO}$ –indicating a prominent enhancement of the gyrotropic activity– develops in the vicinity of $T_C$, only for wavelengths in the interval $\lambda \approx 410 - 475$ nm, whereas for longer wavelengths ($\lambda \approx 600 - 700$ nm) such anomaly disappears. For a comprehensive overview of the interrelation between magneto-optics and polarons, we mapped out the values of $\Im(\delta_K)_{MR}$ and the area under the bump in $\Im(\delta_K)_{MO}$ (shaded areas in Figure 3a) as a function of wavelength and temperature (Figures 3b and 3c). In the MR map (Figure 3b) two prominent areas with the highest $\Im(\delta_K)_{MR}$ stand out, pointing at the abovementioned *SF* and *SP* spectral regions. In contrast, a single peak emerges in the MO map (Figure 3c). Figures 3b and 3c indicate that photons in spectral region *SF* lead to gyrotropic enhancement, while *SP* photons are inactive for that matter.

We endeavored to find alternative optical arrangements in which the MO increase could be larger. For that purpose, we measured the Kerr ellipticity $\varepsilon$ in polar configuration, which usually exhibits larger gyrotropic effects [10,11]. As shown in Figure 4a and inset of Figure 4b, our expectations paid off: a salient bulging protuberance emerged around $T_C$ in the temperature dependence of $\varepsilon$ measured at $\lambda = 402$ nm –i.e., inside spectral region *SF*–. Instead, the anomaly vanishes progressively for wavelengths approaching the spectral region *SP* ($\lambda = 632$ nm, $700$ nm, Figure 4b). For the sake of assessing quantitatively the MO enhancement, we plotted the ratio $\frac{\varepsilon(\lambda=402\,nm)}{\varepsilon(\lambda=700\,nm)}$ for different values of the magnetic field (Figure 4c). One would expect that more polarons survive at lower fields as they are suppressed by magnetic fields; hence the MO increase is expected to become larger as the applied field is reduced. This prediction is nicely borne out by experiments: data in Figure 4a shows that the anomaly becomes more prominent as the field is reduced; secondly, as shown in Figure 4c, the gyrotropic response hikes upward of a staggering 50-fold increase with respect to the background value, underpinning the huge polaron-induced gyrotropic response.

Why the gyrotropic response is increased *only* when the spin is changed during the photoinduced transition? Here we provide a model that explains this observation; the full development is found in the Supplementary Information. The answer lies in the nontrivial interplay between (i) spin-orbit coupling, which induces spin-mixed states that allow describing spin-flips; (ii) the intersite Mn-Mn polaron hopping that allows electronic $e_g$-$e_g$ transitions that are forbidden by parity; and (iii) the Jahn-Teller interaction that enhances the effect of the spin-orbit coupling. To illustrate these points, consider the movement of a spin-up (↑) electron associated to a polaron that is transferred between $e_g$ sites (Fig. 5a). The electron starting on a $d_{z2}$ orbital ($|z_1^2\rangle$) at site 1 moves by jumping preferentially into the neighboring $|z_2^2\rangle$ state at site 2, being the final spin either parallel (↑) or antiparallel (↓). Ignoring momentarily the spin orbit coupling, we find that the only nonzero matrix element is $\langle z_1^2,\uparrow|\vec{r}|z_2^2,\uparrow\rangle$, where $\vec{r}=(x,y,z)$ is the position operator. Thus, in the latter expression, the spin is preserved during the transition. However, if we consider spin-orbit coupling within a center (Fig. 5b and Supplementary Information), the initial state turns into a spin-mixed state $|\tilde{z}_1^2,\uparrow\rangle = |z_1^2,\uparrow\rangle + \sqrt{3}\zeta(|xz_1,\downarrow\rangle + i|yz_1,\downarrow\rangle)/2(\Delta - 2E_{JT})$, where $\Delta$ is the octahedral crystal-field splitting and $E_{JT}$ is the Jahn-Teller energy, both parameters describing the level splitting associated to the distortion of the $MnO_6$ octahedron that enhances the effect of the spin-orbit coupling (see Fig. 5). We can use these states to calculate the probabilities of spin-preserving ($\wp_{SP}^{\uparrow\uparrow}$, $\wp_{SP}^{\downarrow\downarrow}$) and spin-flipping transitions ($\wp_{SF}^{\uparrow\downarrow}$, $\wp_{SF}^{\downarrow\uparrow}$) as follows (see Supplementary Information):

$$\wp_{SP}^{\uparrow\uparrow} = \wp_{SP}^{\downarrow\downarrow} \propto |\langle z_1^2|z|z_2^2\rangle|^2 \qquad \text{(Equation 1)}$$

$$\wp_{SF}^{\uparrow\downarrow} \propto \left(\frac{\zeta}{\Delta-E_{JT}}\right)^2 |\langle xz_1|x|z_2^2\rangle|^2 (1+\sin\varphi) \qquad \text{(Equation 2)}$$

$$\wp_{SF}^{\downarrow\uparrow} \propto \left(\frac{\zeta}{\Delta - E_{JT}}\right)^2 \left|\langle xz_1|x|z_2^2\rangle\right|^2 (1 - \sin\varphi) \qquad \text{(Equation 3)}$$

where $\varphi$ defines the polarization state of light, and $\varphi = \pi/2$ and $\varphi = -\pi/2$ correspond to left- and right- circularly polarized light, respectively. Equation 1 indicates that *SP* transitions are *insensitive* to the polarization of light. However, the probability for transitions in the spectral region *SF* (Equations 2 and 3) is *sensitive* to both the *polarization of light* and the *spin* of the initial electron. In particular, when the initial spin is up (down) the system absorbs left-handed (right-handed) polarized light. This clearly establishes the link between the gyrotropic response in the spectral region *SF* and the spin-flips in polaron hopping. These effects are enhanced when the Jahn-Teller distortion becomes larger. This is precisely what happens in narrow-band manganites when $T_C$ is approached from below, in which electrons become more localized but still retain some mobility to hop.

The aforesaid uncovering of a huge polaron-induced gyrotropic response is fascinating for a number of reasons, which open up enticing perspectives: (i) Beyond approaches based on photonic-band effects [26,27] or plasmonic resonances [28,29], an unreported mechanism for enhanced gyrotropic effects comes to light; (ii) The phenomenon has been observed in the vicinity of room temperature; rising the effect to higher temperatures may be envisioned by materials engineering, e.g. by partial substitution of Ca by Sr or Ba [30,31], which are known to increase the Curie temperature in manganites; and (iii) polarons may be rather sensitive to strains, so that their emergence might be adjusted eventually by electric fields through the piezoelectric effect. If successful, the latter prospect might enable a tremendous modulation of the magneto-optic activity by electric fields, thus opening novel avenues for magnetoelectric coupling beyond the conventional modulation of magnetization.

This work was supported by the Spanish Government through projects MAT2014-56063-C2-1-R, FIS2012-37549-C05-04, MAT2012-37776 and the Severo Ochoa SEV-2015-0496 grant and the Generalitat de Catalunya (2014 SGR 734 project). BC acknowledges his grant FPI BES-2012-059023, RC acknowledges his fellowship from CNPq – Brazil and P.G.F. recognizes support from Ramón and Cajal fellowship RYC-2013-12515.

FIG. 1. (a) Magneto-optical (MO) effects involve changes in the polarization state of light, causing a gyrotropic response. In the presence of a magnetic state (M), polarized light acquires an orthogonal polarization state after reflection (in the case illustrated here, s-polarization –normal to the plane of incidence– adds to the initial p-polarized state –in the plane of incidence–. In contrast, magnetoreflectance (MR) is nongyrotropic and is caused by changes in the intensity of light reflected from magnetized surfaces, without mixing its polarization. (b) Both MO and MR can be detected simultaneously in hysteretic loops of the transverse magneto-optical effect, which is caused by the modulation of p-polarized light reflected from surfaces. (c) The temperature dependence of MO and MR mimic the behavior of magnetism and colossal magnetoresistance, respectively.

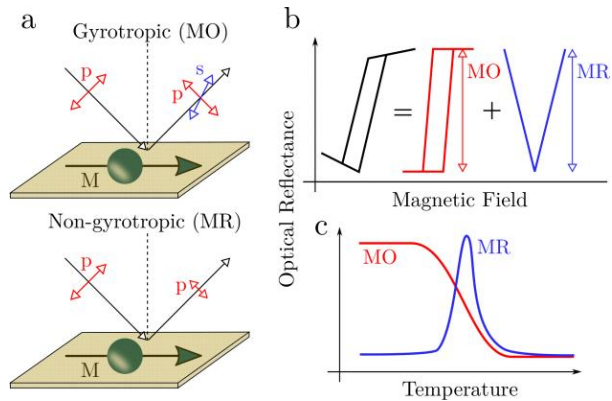

FIG. 2. (a) The different parity relation of the transverse $\Im(\delta_K)$ hysteresis loops measured at $T \approx 256$ K and $\lambda = 475$ nm is used to disentangle the MO and MR curves. (b) The spectral responses of MR and MO are plotted by presenting as a function of wavelength the values of $\Im(\delta_K)_{MO}$ and $\Im(\delta_K)_{MR}$ measured at the highest applied field ($H \sim 15$ kOe). (c) The spectral response of the dielectric function $\tilde{\varepsilon} = \varepsilon_1 + i\varepsilon_2$ is shown. The wavelength dependence of $\Im(\delta_K)_{MO}$, $\Im(\delta_K)_{MR}$ and $\tilde{\varepsilon}$ can be explained in terms of two different photoinduced polaron hopping processes, depicted in (d). For wavelengths in spin-preserving spectral region *SP*, electrons localized in $e_g$ states of $Mn^{3+}$ jump into an empty state of a neighboring $Mn^{4+}$ site, with their spin unchanged. Contrarily, the spins are flipped in transitions induced by higher energy photons in region *SF*.

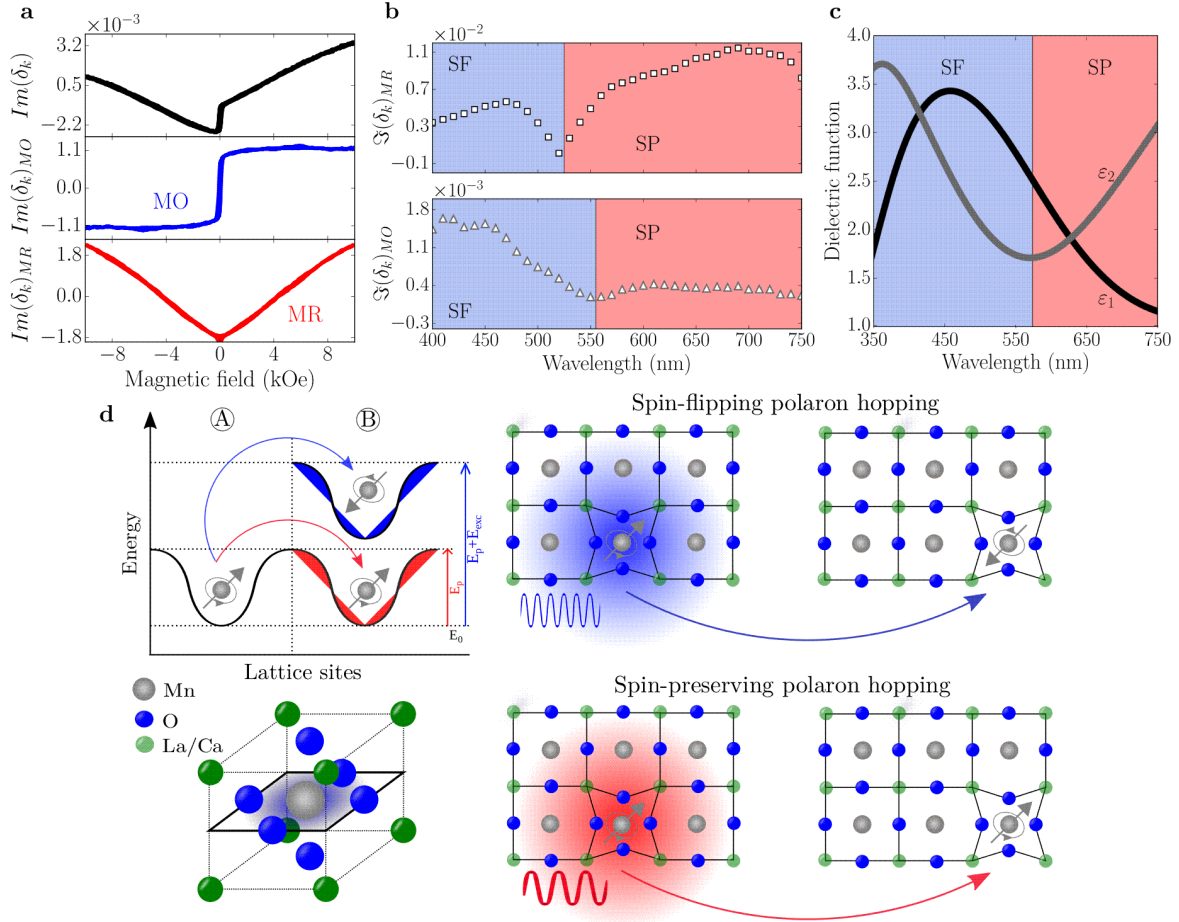

FIG. 3. (a) The temperature dependence of the gyrotropic $\Im(\delta_K)_{MO}$ response measured at the highest applied field ($H \sim 1.5$ kOe) is shown. The areas shaded under the bumps close to $T_C$ are used to estimate the enhancement of the MO signal. In (b), the MR response is mapped as a function of wavelength and temperature. Peaks corresponding to photoinduced small polaron hopping are identified as spin-preserving (*SP*) as well as spin-flipping (*SF*). In (c) we chart as a function of wavelength and temperature the gyrotropic enhancement estimated from the shaded areas in (a).

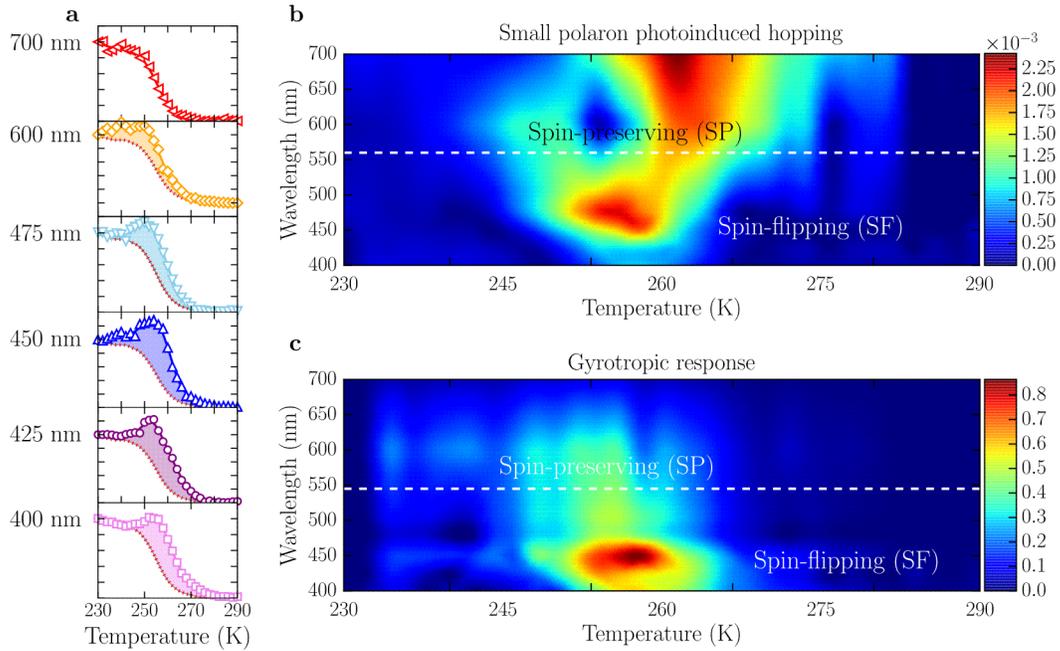

FIG. 4. The temperature dependence of the ellipticity $\varepsilon$ measured in polar configuration at $\lambda = 402$ nm is shown in (a) for a set of applied magnetic fields. A prominent bulge is observed around $T_C$, signaling an outstanding gyrotropic enhancement, which is more visible for decreasing magnetic fields. The evolution of the gyrotropic response is made clear in (b) by selecting three wavelengths. We see that a large increase of $\varepsilon$ around $T_C$ develops at $\lambda = 402$ nm (well within the spin-flipping spectral region *SF* defined in Figures 2b, 2c and 3b, 3c), whereas it is suppressed for longer wavelengths in the spin-preserving *SP* region ($\lambda = 632$ and $700$ nm). The inset in (b) shows the loops measured at three different temperatures around $T_C$. Panel (c) displays the ratio between the gyrotropic responses measured at $\lambda = 402$ nm and $\lambda = 700$ nm, and gives an accurate estimation of the gyrotropic enhancement for a manifold of applied fields. Enhancements upward of $\cong 50$-fold are observed at low fields, hence epitomizing the hugely increased gyrotropic response due to spin-flipping polarons.

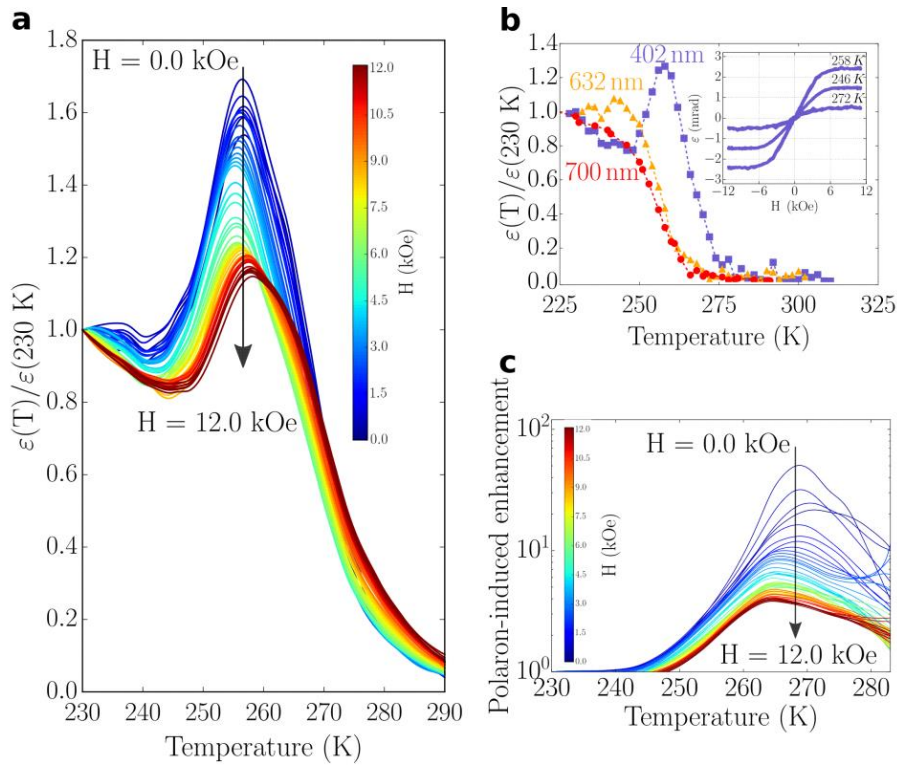

FIG. 5. (a) Scheme of the main process describing the polaron hopping in $La_{2/3}Ca_{1/3}MnO_3$ close to the magnetic transition. A spin-up self-trapped electron moves from the $z^2$ orbital in an elongated $Mn^{III}O_6$ complex into the $z^2$ orbital of a $Mn^{IV}O_6$ complex that can have the same spin (blue box) or spin down (green box). (b) Local $d$-levels in $Mn^{III}$ and $Mn^{IV}$ complexes where the effect of the Jahn-Teller distortion, $Q_{JT}$, is seen, effectively reducing the energy spacing in tetragonal symmetry between the $z^2$ levels and those of $xz$ and $yz$ states. This distortion helps the mixing of occupied and unoccupied levels due to the spin-orbit coupling (shown in purple). (c) Electrons with a particular spin-polarization can only absorb circular polarized light that rotates in a specific direction to invert the spin upon hopping.

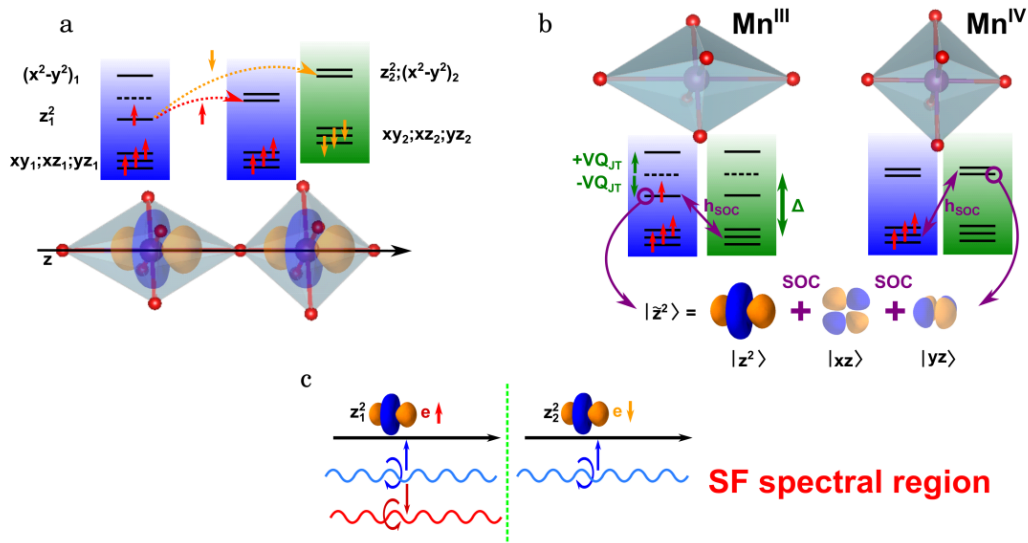

REFERENCES
[1] L.D. Landau, Phys. Z. Sowjetunion 3, 644 (1933).
[2] J.T. Devreese in Part I. Optical Properties of Few and Many Fröhlich Polarons from 3D to 0D. Polarons in advanced materials, edited by A.S. Alexandrov (Springer, 2007).
[3] H. Fröhlich, Adv. Phys. 3, 325 (1954).
[4] A.S. Alexandrov, and N. Mott, in Polarons and Bipolarons (World Scientific Singapore 1995).
[5] A.M. Stoneham et al. J. Phys.: Condens. Matter 19, 255208 (2007).
[6] T. Holstein, Ann. Phys. 8, 325 (1959).
[7] T. Holstein, Ann. Phys. 8, 343 (1959).
[8] G.B. Arnold, and T. Holstein, Phys. Rev. Lett. 33, 1547 (1974).
[9] F.M. Peeters, and J.T. Devreese, Phys. Rev. B 34, 7246 (1986).
[10] A. Zvezdin, and V. Kotov, in Modern Magnetooptics and Magnetooptical Materials (Taylor & Francis: London, 1997).
[11] V. Antonov, B. Harmon, and A. Yaresko, in Electronic Structure and Magneto-Optical Properties of Solids (Springer: New York, 2004).
[12] A.J. Millis, P.B. Littlewood, and B.I. Shraiman, Phys. Rev. Lett. 74, 5144 (1995).
[13] M.B. Salamon, and M. Jaime, Rev. of Mod. Phys. 73, 583-628 (2001).
[14] J.M. Caicedo, M.C. Dekker, K. Dörr, J. Fontcuberta, and G. Herranz, Phys. Rev. B 82, 140410(R) (2010).
[15] J.M. Caicedo, J. Fontcuberta, G. Herranz, Phys. Rev. B 89, 045121 (2014).
[16] M. Quijada, et al., Phys. Rev. B 58, 16093 (1998).
[17] J.H. Jung et al., Phys. Rev. B 55, 15489 (1997).
[18] Y. Okimoto et al., Phys. Rev. Lett. 75, 109 (1995).
[19] J. Fontcuberta et al., Phys. Rev. Lett. 76, 1122 (1996).
[20] H.Y. Hwang, S.-W. Cheong, P.G. Radaelli, M. Marezio, and B. Batlogg, Phys. Rev. Lett. 75, 914 (1995).
[21] A.S. Alexandrov, and A.M. Bratkovsky, J. Appl. Phys. 87, 5016 (2000).
[22] K. Postava et al., J. Eur. Opt. Soc. Rapid Publ. 1, 06017 (2006).
[23] S. Kaplan et al., Phys. Rev. Lett. 77, 2081 (1996).
[24] J.H. Jung, K.H. Kim, T.W. Noh, E.J. Choi, and J. Yu, Phys. Rev. B 57, R11043 (1998).
[25] A.S. Moskvin, A.A. Makhnev, L.V. Nomerovannaya, N.N. Loshkareva, and A.M. Balbashov, Phys. Rev. B 82, 035106 (2010).
[26] J.F. Galisteo-López et al., Adv. Mater. 23, 30 (2011).
[27] E. Yablonovitch, Nature 461, 744 (2009).
[28] V.V. Temnov et al., Nat. Photonics 4, 107 (2010).
[29] N. Maccaferri, et al., Nat. Commun. 6, 6150 (2015).
[30] F. Rivadulla et al., Phys. Rev. Lett. 96, 016402 (2006).
[31] J.M.D. Coey, M. Viret, and S. von Molnár, Adv. Phys. 48, 167-293 (1999).

# Supplementary information:

# Giant Optical Polarization Rotation Induced by Spin-Orbit Coupling in Polarons

Blai Casals[1], Rafael Cichelero[1], Pablo García Fernández[2], Javier Junquera[2], David Pesquera[1], Mariano Campoy-Quiles[1], Ingrid C. Infante[1], Florencio Sánchez[1], Josep Fontcuberta[1], Gervasi Herranz[1]

[1]Institut de Ciència de Materials de Barcelona (ICMAB-CSIC), Campus de la UAB, 08193 Bellaterra, Catalonia, Spain

[2]Departamento de Ciencias de la Tierra y Física de la Materia Condensada, Universidad de Cantabria, Cantabria Campus Internacional, Avenida de los Castros s/n, 39005 Santander, Spain

### A. Selectivity in the absorption of polarized light in photoinduced polaron transitions

In the metallic and ferromagnetic state of $La_{2/3}Ca_{1/3}MnO_3$ $e_g$ electrons have relatively large mean free paths before they are scattered. As the temperature is raised towards the magnetic transition this length is reduced due to the higher probability of finding a Mn site with opposite spin that acts as a scattering site. As soon as the transition temperature is reached, the optical activity drops abruptly since the $e_g$ electrons are essentially localized in a single Mn ion and intrasite $d$-$d$ transitions are forbidden by parity.

In order to understand qualitatively the origin of the magneto-optical properties close to the ferromagnetic-paramagnetic transition in $La_{2/3}Ca_{1/3}MnO_3$ we will consider a model in which an electron is moving in the conduction band of this perovskite, which has a strong $Mn(e_g)$ character, and has to hop into a $Mn^{IV}$ ion with an empty $e_g$ shell (see Fig. 5a of the main text). An important fact to take into account is that when an electron is localized in the $e_g$ shell of an octahedral Mn complex the resulting degenerate state (see Fig. 5b of the main text), triggers a geometrical instability in the form of the Jahn-Teller effect. As a consequence, the complex becomes elongated and electrons occupy preferentially orbitals with $3d_{3z^2-r^2}$ symmetry. In $La_{2/3}Ca_{1/3}MnO_3$ this distortion can be experimentally observed by EXAFS and is found to increase as the temperature approaches the magnetic transition [1]. This can be understood in terms of the interactions that cause the Jahn-Teller effect, which are proportional to the occupation of each complex and are thus enhanced by electron localization. This effect allows the electrons to self-trap and form polarons. In larger bandwidth perovskites like La$_{2/3}$Sr$_{1/3}$MnO$_3$ the localization is much smaller and electron self-trapping is impeded.

### A.1 Mechanism of the photoinduced hopping

To understand how the polaron moves between complexes we will take a two-site model where the electron is initially localized in the mainly $3d_{3z^2-r^2}$ orbital ($|z_1^2\rangle$) of one manganese ion with nominal charge +3 (Mn$^{III}$) and moves towards a contiguous Mn$^{IV}$ that presents an empty $e_g$ shell. First, we will establish that the most probable mechanism for this hopping mechanism involves the direct charge transfer from $|z_1^2\rangle$ to the mainly $3d_{3z^2-r^2}$ orbital in the second centre ($|z_2^2\rangle$). To do so we will consider the movement of an electron in a generic two-center tight-binding model where each site has a coupling to local vibrations $Q_i$. The energy of this system is:

$$E = \frac{1}{2}KQ_1^2 + \frac{1}{2}KQ_2^2 + \varepsilon_1 \qquad \text{(Equation S1)}$$

where $K$ is the local force constant associated to the distortions and $\varepsilon_1$ is the energy of the electron, obtained as the lowest eigenvalue of the Hamiltonian matrix. In order to build this matrix we will consider that the electronic state on each site can be described by a single Wannier function. Defining $V$ as the linear electron-lattice coupling constant (equivalent to the Jahn-Teller linear coupling in systems with degenerate electronic levels) and $\gamma$ as the tight-binding constant associated to the interaction between the two Wannier functions, we obtain the following interaction matrix:

$$h = \begin{pmatrix} -VQ_1 & \gamma \\ \gamma & -VQ_2 \end{pmatrix} \qquad \text{(Equation S2)}$$

Using Equation S1 we can now calculate the polaron's binding energy,

$$E_b \approx \frac{V^2}{2K} \qquad \text{(Equation S3)}$$

and the adiabatic barrier for the charge transfer of the electron from site 1 to site 2 at first order in $\gamma$:

$$B \approx \frac{V^2}{4K} - |\gamma| + \cdots \qquad \text{(Equation S4)}$$

Hence, we see that the barrier reduces as the interaction $\gamma$ becomes larger. We can now determine the smallest barrier for polaron hopping by using standard Koster-Slater constants to find the largest hopping constant between the $|z_1^2\rangle$ orbitals and the *d*-orbitals of the surrounding Mn ions. As already anticipated, the result is that the most probable hop is to the mainly $3d_{3z^2-r^2}$ orbitals of the complexes along the *z*-direction.

While we developed this simple model for a generic non-degenerate case, its extension to the two Jahn-Teller sites necessary for its application in the manganites follows the same pattern where the binding energy can be directly identified with the Jahn-Teller energy (this is an approximation given that other vibrational modes like the octahedral breathing and rotation modes also participate in the stabilization of the polaron).

**A.2 Optical activity associated to the photoinduced polaron hopping**

It is well known that due to the even character of the 3*d*-levels of an octahedral complex, *d*-*d* transitions in a single transition site are forbidden by parity. However, in the two-site model a new dipole-allowed transition is opened. In particular, if we consider that the electron is initially spin-up ($S = \uparrow$) we find that the only non-zero transition dipole matrix element is $\langle z_1^2, S|z|z_2^2, S\rangle$ associated to the spin-conserving transfer of the electron from the mainly $3d_{3z^2-r^2}$ orbital in one centre to the other. Thus it is clear that, at this level of theory, no magneto-optical effect can be found associated to polaron hopping.

In order to find these effects we will include the effect of the spin-orbit coupling between the mainly $3d_{3z^2-r^2}$ levels with other *d*-levels in the same site. We take the usual expression for the one-site spin-orbit operator,

$$\hat{h}_{SOC} = \zeta(r)\vec{l}\cdot\vec{s} \qquad \text{(Equation S5)}$$

where $\zeta(r)$ is the radial part of the spin-orbit interaction and $\vec{l}$ and $\vec{s}$ are the orbital and spin angular momentum operators. Given that the $|z_1^2\rangle$ state has zero orbital angular momentum – it has the same symmetry as the $Y_2^0$ spherical harmonic – it can only couple to functions with ±1 orbital angular momentum, i.e. $|xz\rangle$ and $|yz\rangle$, through the $\hat{l}_x$ and $\hat{l}_y$ operators. Given that the spin-orbit coupling interaction is represented here as a one-electron operator, the energy of the system can only become lower when occupied and unoccupied orbitals are mixed (see Fig. 5b of the main text). We find that, at first perturbation order, the initial state with the electron spin-up is corrected to

$$|\tilde{z}_1^2 \uparrow\rangle \approx |z_1^2 \uparrow\rangle + \frac{\sqrt{3}}{2}\frac{\zeta}{\Delta - V Q_{JT}}(|xz_1 \downarrow\rangle + i|yz_1 \downarrow\rangle) + \cdots \qquad \text{(Equation S6)}$$

while the corresponding spin-down case is corrected to

$$|\tilde{z}_1^2 \downarrow\rangle \approx |z_1^2 \downarrow\rangle + \frac{\sqrt{3}}{2}\frac{\zeta}{\Delta - V Q_{JT}}(|xz_1 \downarrow\rangle - i|yz_1 \uparrow\rangle) + \cdots \qquad \text{(Equation S7)}$$

It is important to note that the second term in Equations S6 and S7 corresponds to functions with well-defined orbital angular momentum ($l_z = \pm 1$ for Equations S6 and S7, respectively), so the above equations can be rephrased into

$$|\tilde{z}_1^2 \uparrow\rangle \approx |z_1^2 \uparrow\rangle + \frac{\sqrt{3}}{2}\frac{\zeta}{\Delta - V Q_{JT}}|l_{+1} \downarrow\rangle + \cdots \qquad \text{(Equation S8)}$$

$$|\tilde{z}_1^2 \downarrow\rangle \approx |z_1^2 \downarrow\rangle + \frac{\sqrt{3}}{2}\frac{\zeta}{\Delta - V Q_{JT}}|l_{-1} \downarrow\rangle + \cdots \qquad \text{(Equation S9)}$$

The well-defined angular momenta in the spin-orbit-corrected states of Equations S8 and S9 influence decisively on the sensitivity of the states with spin-up and spin-down to absorbing right and left-handed circularly polarized light.

For the second site, which initially contains the Mn$^{IV}$ ion with the empty $e_g$ shell, we find that the first-order spin-orbit correction to the orbital receiving the electron is null both when the global spin of the ion is up,

$$|\tilde{z}_2^2 \uparrow\rangle = |z_2^2 \uparrow\rangle \qquad \text{(Equation S10)}$$

or down,

$$|\tilde{z}_2^2 \downarrow\rangle = |z_2^2 \downarrow\rangle \qquad \text{(Equation S11)}$$

In the above Equations S6-S9, $\zeta$ is the covalency-corrected spin-orbit coupling constant of Mn, $\Delta$ is the so-called $10Dq$ crystal splitting for the Mn$^{III}$ ion, $V$ is the linear Jahn-Teller coupling constant and $Q_{JT}$ is the Jahn-Teller distortion (see Fig. 5 of the main text for an illustration of the meaning of these magnitudes).

Now we can consider the optical properties associated to two photoinduced hopping processes denoted in the main text as spin-preserving (*SP*) and spin-flipping (*SF*). We are going to study each situation considering that the initial electron is spin-polarized up or down to find the differences in interaction with polarized light. In the case of *SP*, the spin is conserved during the charge transfer so that,

$$Mn^{III}(\uparrow) + Mn^{IV}(\uparrow) \rightarrow Mn^{IV}(\uparrow) + Mn^{III}(\uparrow) \quad \text{(Equation S12)}$$

$$Mn^{III}(\downarrow) + Mn^{IV}(\downarrow) \rightarrow Mn^{IV}(\downarrow) + Mn^{III}(\downarrow) \quad \text{(Equation S13)}$$

In spectral region *SF* the spin is reversed so the two processes to consider are:

$$Mn^{III}(\uparrow) + Mn^{IV}(\downarrow) \rightarrow Mn^{IV}(\uparrow) + Mn^{III}(\downarrow) \quad \text{(Equation S14)}$$

$$Mn^{III}(\downarrow) + Mn^{IV}(\uparrow) \rightarrow Mn^{IV}(\downarrow) + Mn^{III}(\uparrow) \quad \text{(Equation S15)}$$

To qualitatively estimate the optical activity we will use Fermi's Golden rule that says that the transition probability is proportional to the squared modulus of the transition matrix element. We will consider here the radiation-electron interaction Hamiltonian of the form,

$$\hat{h}_E = -\vec{E}(t)\cdot\vec{r} \quad \text{(Equation S16)}$$

Thus, we need to evaluate the corresponding dipole transition matrix elements, $\langle \tilde{z}_1^2, S_1 | \vec{r} | \tilde{z}_2^2, S_2 \rangle$, for each case (*SP* or *SF*) and starting with spin-up or spin-down. For the case *SP* and using Equations S6-S7 and S8-S9 we find that the only non-null matrix elements are,

$$\langle \tilde{z}_1^2, \uparrow | z | \tilde{z}_2^2, \uparrow \rangle = \langle \tilde{z}_1^2, \downarrow | z | \tilde{z}_2^2, \downarrow \rangle = \langle z_1^2 | z | z_2^2 \rangle \quad \text{(Equation S17)}$$

while for case *SF* we get that the non-null elements are,

$$\langle \tilde{z}_1^2 \uparrow |x| \tilde{z}_2^2 \downarrow \rangle \approx \frac{\sqrt{3}}{2} \frac{\zeta}{\Delta - V_{QJT}} \langle xz_1 |x| z_2^2 \rangle \qquad \text{(Equation S18)}$$

$$\langle \tilde{z}_1^2 \uparrow |y| \tilde{z}_2^2 \downarrow \rangle \approx -i \frac{\sqrt{3}}{2} \frac{\zeta}{\Delta - V_{QJT}} \langle yz_1 |y| z_2^2 \rangle \qquad \text{(Equation S19)}$$

$$\langle \tilde{z}_1^2 \downarrow |x| \tilde{z}_2^2 \uparrow \rangle \approx \frac{\sqrt{3}}{2} \frac{\zeta}{\Delta - V_{QJT}} \langle xz_1 |x| z_2^2 \rangle \qquad \text{(Equation S20)}$$

$$\langle \tilde{z}_1^2 \downarrow |y| \tilde{z}_2^2 \uparrow \rangle \approx i \frac{\sqrt{3}}{2} \frac{\zeta}{\Delta - V_{QJT}} \langle yz_1 |y| z_2^2 \rangle \qquad \text{(Equation S21)}$$

where $(x, y, z)$ designate the projections of the vector $\vec{r}$ along the three directions in space. We will now consider an experiment in polar configuration (the main magnetic axis is along direction $z$) with light propagating also along $z$ and a polarization described by a phase factor $\varphi$ between the electric field components along the directions $x$ and $y$.

Let us focus first on case *SP* (spin-conserving electron transfer). Given that no electric field component is along $z$, Equation S17 tells us that we would not expect any optical activity along this direction. However, the matrix element $\langle z_1^2 |z| z_2^2 \rangle$ is equal to $\langle x_1^2 |x| x_2^2 \rangle$ and $\langle y_1^2 |y| y_2^2 \rangle$, which are their equivalents along the $x$ and $y$ directions. Note that $(|x_i^2 \rangle$ and $|y_i^2 \rangle$ represent, respectively, $3d_{3x^2-r^2}$ and $3d_{3y^2-r^2}$ orbitals. Thus, the transition probability for case *SP* (processes described by Equations S10 and S11) are:

$$\wp_{SP}^{\uparrow\uparrow} \propto \frac{1}{3} |\langle \tilde{x}_1^2, \uparrow |x| \tilde{x}_2^2, \uparrow \rangle|^2 + \frac{1}{3} |\langle \tilde{y}_1^2, \uparrow |y| \tilde{y}_2^2, \uparrow \rangle|^2 = \frac{2}{3} |\langle z_1^2 |z| z_2^2 \rangle|^2 \qquad \text{(Equation S22)}$$

$$\wp_{SP}^{\downarrow\downarrow} \propto \frac{1}{3} |\langle \tilde{x}_1^2, \downarrow |x| \tilde{x}_2^2, \downarrow \rangle|^2 + \frac{1}{3} |\langle \tilde{y}_1^2, \downarrow |y| \tilde{y}_2^2, \downarrow \rangle|^2 = \frac{2}{3} |\langle z_1^2 |z| z_2^2 \rangle|^2 \qquad \text{(Equation S23)}$$

Thus, we see that for the *SP* case –independently of the magnetic polarization of the sample– the optical signal is insensitive to the polarization of light, i.e. *the transitions in spectral region SP do not present magneto-optical activity*.

Operating in a similar fashion for the spin-flip processes (case *SF*) described in Equations S14 and S15 we get,

$$\wp_{SF}^{\uparrow\downarrow} \propto |\langle \tilde{z}_1^2, \uparrow | \vec{r} | \tilde{z}_2^2, \downarrow \rangle|^2 = 3 \left( \frac{\zeta}{\Delta - vQ_{JT}} \right)^2 |\langle xz_1 | x | z_2^2 \rangle|^2 (1 + \sin \varphi) \qquad \text{(Equation S24)}$$

$$\wp_{SF}^{\downarrow\uparrow} \propto |\langle \tilde{z}_1^2, \downarrow | \vec{r} | \tilde{z}_2^2, \uparrow \rangle|^2 = 3 \left( \frac{\zeta}{\Delta - vQ_{JT}} \right)^2 |\langle xz_1 | x | z_2^2 \rangle|^2 (1 - \sin \varphi) \qquad \text{(Equation S25)}$$

where we plainly see that systems with starting spin-up or spin-down electrons interact with polarized light in a clearly different way. In particular, we observe that each case (either Equation S24 or S25) is sensitive, respectively, to just one circularly polarized state (either $\varphi = +\pi/2$ or $\varphi = -\pi/2$) and is completely insensitive to the other. Thus, we can conclude that *Equations S21 and S22 describe the magneto-optical effect associated to the polaron hopping in the spectral region SF (when the spin is reversed during the transition)*.

Looking at Equations S8 and S9 we can easily understand why these transitions are sensitive to polarized light. In our set-up for the polaron movement, we have an initial angular momentum $J_z=1/2$ and an axial symmetry. Thus, processes must conserve this total angular momentum and simple spin-reversal is forbidden as it alters this quantity. However, under polarized light the necessary orbital angular momentum can be gained through the radiation, yielding the magneto-optical effects associated to spectral region *SF* (see Fig. 5c in the main text).

Moreover, using Equations S24 and S25 we can now understand the enhancement of the gyrotropic activity as the system gets closer to the transition temperature. As previously remarked, the average Jahn-Teller distortion $Q_{JT}$ gets larger as the temperature rises towards $T_C$. This means that as electrons get more localized approaching the magnetic transition the denominator in Equations S24-S25 gets smaller and the transition probability increases. Thus, the electron-lattice coupling in these systems plays an important role in these perovskites enhancing the effect of the spin-orbit coupling.

### B. Description of the analysis of ellipsometry measurements

Variable angle spectroscopic ellipsometry was measured using a rotating polarizer GES5E ellipsometer with CCD detection from SOPRA (SEMILAB). In this experiment we measured an 80 nm thick $La_{2/3}Ca_{1/3}MnO_3$ thin film grown on top of a $SrTiO_3$ substrate. Four incident angles were measured for each sample (60, 65, 70 and 75 degrees) and the spectra were detected for photon energies in the range between 1.2184 eV and 5.5299 eV.

For the SrTiO$_3$ crystal, the data was analysed using a Tauc-Lorentz model assuming an infinite thickness [2] [2]. This model allows defining a bandgap and thus correcting for the non-zero absorption below the gap often observed when using the Lorentz model. Two transitions were included. The resulting dielectric function is shown in figure S1. Note that the fit to the ellipsometric data did not improve when including surface roughness or anisotropy.

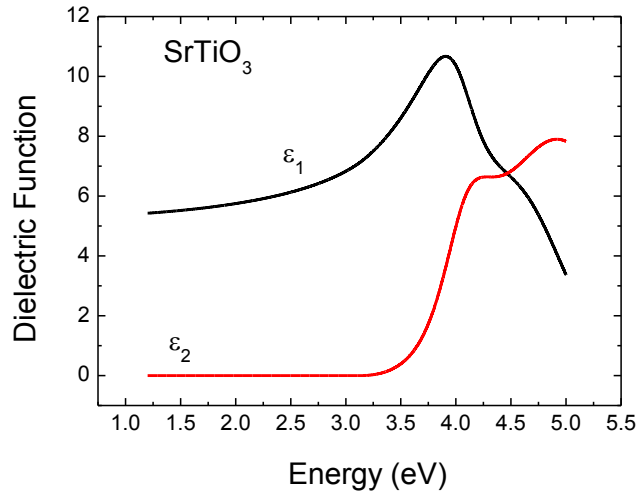

**Figure S1**. *Dependence of the real ($\varepsilon_1$) and imaginary ($\varepsilon_2$) parts of the dielectric function as a function of the photon energy, measured in a SrTiO$_3$ substrate.*

For the La$_{2/3}$Ca$_{1/3}$MnO$_3$ thin film, we employed two Gaussian peaks, to account for the two polaronic transitions, and one Lorentzian peak to account for a UV absorption tail. The thickness was fixed at the nominal value of 80 nm since no transparency region was available to deduce the film thickness with certainty. A sensitivity analysis showed that the final results are only slightly sensitive to small changes in film thickness (around 10 nm). Reassuringly, assuming a very crude semi-infinite layer (infinite thickness) resulted in qualitatively similar results, with quantitative differences smaller than 30% as shown in Figure S2. The dielectric function of SrTiO$_3$ described above was used as the substrate in the modelling of the thin film. No improvement in the fit was observed when using anisotropic models. The deduced dielectric function is shown in Figure 2c of the main text. Additionally, we used the Drude-Lorentz model [2] to fit the data and check the sensitivity to the particular model employed, finding very good agreement between the dielectric functions obtained using both models (Figure S2).

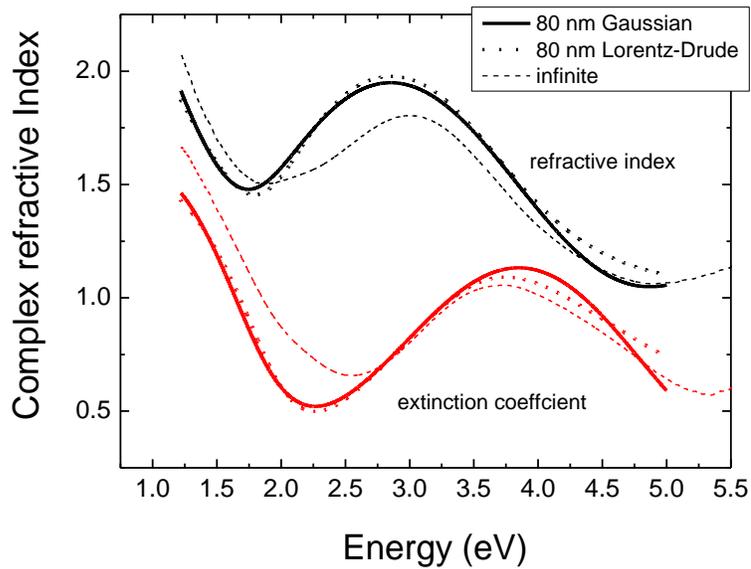

**Figure S2**. *Comparison of three different models for the analysis of the VASE data for the thin $La_{2/3}Ca_{1/3}MnO_3$ film, namely, Gaussian (solid line), Lorentz-Drude (dotted-line) and semi-infinite (dashed line) models. The real and imaginary parts are shown in black and red colour, respectively.*

**C. Temperature dependence of the polaronic MR response and its relation to the enhancement of the gyrotropic response.**

We have measured the optical properties of different $La_{2/3}Ca_{1/3}MnO_3$ thin films with different thicknesses and crystal orientations. The outcome is that the observation of the giant gyrotropic response is possible, provided that the magnetoreflectance (MR) associated with small polarons is observed within a narrow enough range of temperatures around the transition $T_C$. This is clearly revealed by Figure S3, which shows the magneto-optical response (ellipticity) for three samples. The gyrotropic enhancement is outstanding for the 93 nm thick $La_{2/3}Ca_{1/3}MnO_3$ thin film grown on (110)-oriented $SrTiO_3$ substrate, for which the MR response is narrowly distributed around the transition. However, for the thinner (110)-oriented film (thickness 17 nm), the MR response is much weaker and distributed over a broader range of temperatures; as a consequence, the gyrotropic enhancement is not observed. For a 93 nm thick film grown on (001)-oriented $SrTiO_3$ substrate, the MR response is distributed around a broader range of temperatures than that of the 93 nm thick (110)-oriented sample, but the MR response is stronger than that of the (110)-oriented 17 nm thick film. As a consequence, a subtle enhanced gyrotropic response can be observed, although much weaker and broad than the one observed in the 93 nm thick (110)-oriented sample. We can conclude, then, that the distribution of polarons around the transition temperature is extremely critical to observe the phenomenon. Therefore, a sharp metal insulator transition –with a narrow distribution of polarons, particularly peaked around the transition– is required to observe the giant gyrotropic enhancement.

In this regard, it is interesting to note that cationic disorder plays a critical in the transport properties of manganites [1]. One can imagine that disorder may create inhomogeneous strain distributions associated, e.g., to nonuniform cationic distributions or the presence of vacancies that, in turn, may stabilize the presence of polarons over much wider ranges of temperatures than in the bulk crystal, thus weakening the gyrotropic response of polarons. Indeed, there is evidence that these effects are more prominent for (001)-oriented samples than for (110)-oriented films [3]. This would explain why the enhanced gyrotropic response is best observed for $La_{2/3}Ca_{1/3}MnO_3$ thin films grown on $SrTiO_3$ substrates with (110) orientation. On the other hand, it is rather common to find enhanced disorder close to interfaces, which explains why the transition is sharper as the thickness increases [4] and why the polaronic gyrotropic response is suppressed in the 17 nm thick film. Also, the different structural relaxation mechanisms of (001) and (110) oriented films is consistent with sharper transitions for the latter at a given value of thickness [4].

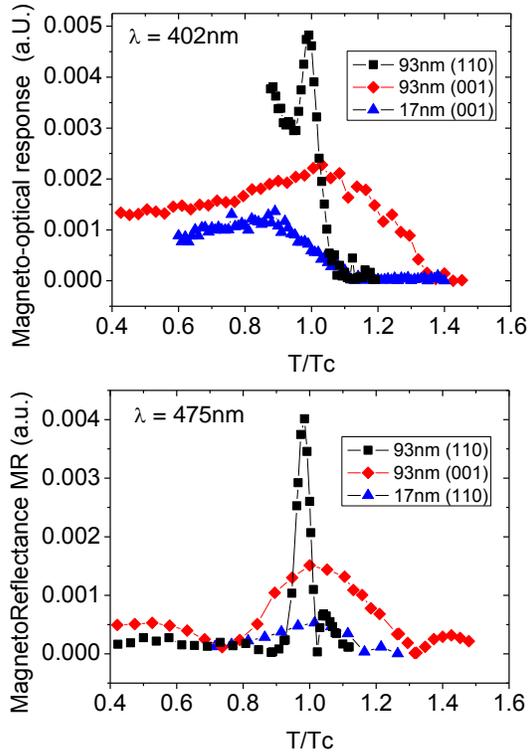

**Figure S3**. Upper panel: *The magneto-optical response (ellipticity) of three different $La_{2/3}Ca_{1/3}MnO_3$ thin films is shown as a function of the temperature normalized to the transition temperature $T_C$.* Lower panel: *the magnetoreflectance (MR) is displayed against the normalized temperature.*

[1] Rodriguez-Martinez, L.M. & Attfield, J.P. Cation disorder and the metal-insulator transition temperature in manganese oxide perovskites Phys. Rev. B 58, 2426-2429 (1998).

[2] Handbook of ellipsometry, ed. by Tompkins, H.G. & Irene, E.A., Springer 2005.

[3] Estradé, S. et al., Cationic and charge segregation in $La_{2/3}Ca_{1/3}MnO_3$ thin films grown on (001) and (110) $SrTiO_3$. Appl. Phys. Lett. 93, 112505 (2008).

[4] Infante, I.C. et al. Elastic and orbital effects on thickness-dependent properties of manganite thin films. Phys. Rev. B 76, 224415 (2007).